\documentclass[aps,prd,twocolumn]{revtex4-1}

\usepackage{aas_macros}
\usepackage{epsfig}
\usepackage{latexsym}
\usepackage{times}

\newcommand{\bed}{\begin{displaymath}}
\newcommand{\eed}{\end{displaymath}}
\newcommand{\bef}{\begin{figure}}
\newcommand{\eef}{\end{figure}}
\newcommand{\bei}{\begin{itemize}}
\newcommand{\eei}{\end{itemize}}
\newcommand{\ben}{\begin{enumerate}}
\newcommand{\een}{\end{enumerate}}
\newcommand{\beq}{\begin{equation}}
\newcommand{\eeq}{\end{equation}}
\newcommand{\ber}{\begin{eqnarray}}
\newcommand{\eer}{\end{eqnarray}}

\newcommand{\msun}{\mbox{{\rm M}$_{\odot}$}}

\newcommand{\gsim}{\raisebox{-0.3ex}{\mbox{$\stackrel{>}{_\sim} \,$}}}

\begin{document}

\title{Comments on ``Strongly magnetized cold degenerate electron gas: Mass-radius relation of the magnetized white dwarf''}
\author{Rajaram Nityananda}
\affiliation{Indian Institute of Science Education and Research, Pune 
             411008, India}
\email[]{rajaram.nityananda@gmail.com,  \\
         now at Azim Premji University, Bangalore}
\author{Sushan Konar}
\affiliation{National Centre for Radio Astronomy - Tata Institute of 
             Fundamental Research, Pune 411007, India}
\email[]{sushan@ncra.tifr.res.in}
\homepage[]{http://www.ncra.tifr.res.in:8081/~sushan/}

\begin{abstract}
The super-massive white dwarf  models proposed by Das \& Mukhopadhyaya
[Phys. Rev. D {\bf 86} 042001 (2012)], based on modifying the equation
of state by a super-strong magnetic  field in the centre, are very far
from  equilibrium because  of the  neglect of  Lorentz forces,  as has
recently been shown by Nityananda \& Konar [arXiv:1306.1625].
\end{abstract}



\maketitle

\date{\today}

Recently, it  has been proposed  that white dwarfs  (stars supported
against   gravity   by   electron   degeneracy  pressure)   can   have
super-Chandrasekhar (2.3  - 2.6~\msun) masses with  radii much smaller
than   previously    thought   to    be   possible   ($\sim$    70   -
600~km)~\cite{das12a,das12b}.  These  models are based  on the altered
equation of state  coming from the quantization of  electron motion in
super-strong ($\gsim 10^{15}$~G) magnetic fields.

The issue  here is a counter-intuitive  feature of such  models -- the
magnetic  pressure  ($P_m$)  is   not  included  in  the  equation  of
hydrostatic equilibrium  even though its  value at the  centre exceeds
the electron pressure ($P_e$) by  nearly two orders of magnitude.  The
justification   given  (in   \cite{das12a}  and \cite{das13c}) is that  
the field is uniform in  the central region of
interest, and therefore there is  no force coming from the gradient of
$P_m$. The field is then presumed  to taper off to much smaller values
at the  surface (consistent  with observations) without  affecting the
analysis which relies solely on electron pressure.

Consider a star of mass $M$, radius $R$ with a magnetic field $B$.  In
an  averaged spherically  symmetric  model (as  in \cite{das12a})  the
equation of hydrostatic  equilibrium, at a radius $r$  inside the star
reads,
\beq
\frac{d P_e}{dr} + \frac{d P_m}{dr} = - \rho(r) g(r), 
\label{eqn01}
\eeq
where $g(r) (=G M(r)/r^2)$  is the radially inward gravitational force
on a unit mass.  Integrating  both sides of Eq.(\ref{eqn01}), from the
centre to the surface of the star, denoted by suffixes $c$ and $s$, we
obtain
\beq
(P_{\rm ec} - P_{\rm es}) + (P_{\rm  mc} - P_{\rm ms}) = \int \rho (r)
g(r) dr.
\label{eqn02}
\eeq
The second bracket  on the right hand side should  be smaller than the
first, if the neglect of  the magnetic contribution to the hydrostatic
equation  is  to  be  valid  everywhere.  But  (equating  the  surface
pressures to zero) the exact opposite is true in these models when the
central  magnetic pressure  becomes larger  than the  pressure  of the
degenerate electrons.

No attempt to taper off a  large, uniform magnetic field in the centre
to  essentially zero  ($B_s/B_c  << 1$)  at  the surface  can avoid  a
gradient of  $P_m$ to be much larger  than that of $P_e$  (at least in
certain   locations),   which  balances   gravity   in  the   proposed
models~\cite{nitya13}.
 
The above argument would be  strictly applicable only if the field was
sufficiently  disordered to  result in  an average  isotropic pressure
within a region smaller than  the scale length over which pressure and
density vary significantly. It has been pointed out~\cite{das13c} that
magnetic  tension, in  an  ordered  field, could  actually  act in  an
opposite  way to magnetic  pressure, and  possibly play  a significant
role  in stabilizing  a  non-spherical configuration  with a  suitably
ordered field.  We have  shown that ordered  fields in  an anisotropic
model  can not  qualitatively change  the conclusions  drawn  from the
averaged  spherical  model~\cite{nitya13},  using  the  property  that
tension  in one direction  is offset  by lateral  pressure in  the two
perpendicular directions.

It should be noted further that the electrons, in the proposed models,
have  Fermi  energies  significantly  above  their  rest  energy  (for
example, $E_{\rm F} \sim 20$ m$_e$ c$^2$).  When the magnetic pressure
is two or more orders  of magnitude greater than the electron pressure
it implies that  the rest energy of the magnetic  field could be close
to  the rest  energy  of the  nuclei,  per electron.   This makes  the
equation of state approach $p \propto \rho$ and points to the need for
a general relativistic treatment  since pressure gravitates in general
relativity.

Another  important  factor  under  such conditions  is  neutronization
(inverse  $\beta$-decay), as  a result  of which  the  electron number
decreases  and the  ionic component  of the  pressure (which  has been
completely neglected so far) begins to become important.  Moreover, at
higher densities  pycnonuclear reactions would  modify the composition
of the matter making it even more difficult for the star to be treated
as an ordinary white dwarf (with standard compositions).  The proposed
formalism does not provide for these energetically favorable processes
either.  We find that while  setting up the modified equation of state
for the electrons, in  presence of quantizing field, due consideration
has not been given to these aspects.

Though some of these concerns regarding the equation of state have now
been  addressed~\cite{chame13,coelh13}, the  basic  concern about  the
equilibrium  of  the  proposed  models  remain.  The  results  from  a
self-consistent   investigation  into   the  structures   of  strongly
magnetized white dwarfs has just become available~\cite{bera14} and it
is  seen  that  while  the  masses  of  such  objects  do  exceed  the
traditional  Chandrasekhar limit  significantly (M  $\sim$ 1.9~\msun),
neither do  the radii  and the density  profiles deviate too  far from
those of  the non-magnetized  white dwarfs nor  are the  maximum field
strengths ($B \sim 10^{14}$~G) very different from those expected from
simple stability arguments~\cite{nitya13}.

In summary, we find that  a fortuitous modification of the equation of
state in presence of strong  magnetic fields to be responsible for the
super-Chandrasekhar     mass      obtained     in     the     proposed
models~\cite{das12a,das12b,das13a,das13b}   with  a  neglect   of  the
magnetic pressure.


\bibliography{adsrefs}

\begin{thebibliography}{9}%
\makeatletter
\providecommand \@ifxundefined [1]{%
 \@ifx{#1\undefined}
}%
\providecommand \@ifnum [1]{%
 \ifnum #1\expandafter \@firstoftwo
 \else \expandafter \@secondoftwo
 \fi
}%
\providecommand \@ifx [1]{%
 \ifx #1\expandafter \@firstoftwo
 \else \expandafter \@secondoftwo
 \fi
}%
\providecommand \natexlab [1]{#1}%
\providecommand \enquote  [1]{``#1''}%
\providecommand \bibnamefont  [1]{#1}%
\providecommand \bibfnamefont [1]{#1}%
\providecommand \citenamefont [1]{#1}%
\providecommand \href@noop [0]{\@secondoftwo}%
\providecommand \href [0]{\begingroup \@sanitize@url \@href}%
\providecommand \@href[1]{\@@startlink{#1}\@@href}%
\providecommand \@@href[1]{\endgroup#1\@@endlink}%
\providecommand \@sanitize@url [0]{\catcode `\\12\catcode `\$12\catcode
  `\&12\catcode `\#12\catcode `\^12\catcode `\_12\catcode `\%12\relax}%
\providecommand \@@startlink[1]{}%
\providecommand \@@endlink[0]{}%
\providecommand \url  [0]{\begingroup\@sanitize@url \@url }%
\providecommand \@url [1]{\endgroup\@href {#1}{\urlprefix }}%
\providecommand \urlprefix  [0]{URL }%
\providecommand \Eprint [0]{\href }%
\providecommand \doibase [0]{http://dx.doi.org/}%
\providecommand \selectlanguage [0]{\@gobble}%
\providecommand \bibinfo  [0]{\@secondoftwo}%
\providecommand \bibfield  [0]{\@secondoftwo}%
\providecommand \translation [1]{[#1]}%
\providecommand \BibitemOpen [0]{}%
\providecommand \bibitemStop [0]{}%
\providecommand \bibitemNoStop [0]{.\EOS\space}%
\providecommand \EOS [0]{\spacefactor3000\relax}%
\providecommand \BibitemShut  [1]{\csname bibitem#1\endcsname}%
\let\auto@bib@innerbib\@empty
\bibitem [{\citenamefont {{Das}}\ and\ \citenamefont
  {{Mukhopadhyay}}(2012{\natexlab{a}})}]{das12a}%
  \BibitemOpen
  \bibfield  {author} {\bibinfo {author} {\bibfnamefont {U.}~\bibnamefont
  {{Das}}}\ and\ \bibinfo {author} {\bibfnamefont {B.}~\bibnamefont
  {{Mukhopadhyay}}},\ }\href {\doibase 10.1103/PhysRevD.86.042001} {\bibfield
  {journal} {\bibinfo  {journal} {\prd}\ }\textbf {\bibinfo {volume} {86}},\
  \bibinfo {eid} {042001} (\bibinfo {year} {2012}{\natexlab{a}})},\ \Eprint
  {http://arxiv.org/abs/1204.1262} {arXiv:1204.1262 [astro-ph.HE]} \BibitemShut
  {NoStop}%
\bibitem [{\citenamefont {{Das}}\ and\ \citenamefont
  {{Mukhopadhyay}}(2012{\natexlab{b}})}]{das12b}%
  \BibitemOpen
  \bibfield  {author} {\bibinfo {author} {\bibfnamefont {U.}~\bibnamefont
  {{Das}}}\ and\ \bibinfo {author} {\bibfnamefont {B.}~\bibnamefont
  {{Mukhopadhyay}}},\ }\href {\doibase 10.1142/S0218271812420011} {\bibfield
  {journal} {\bibinfo  {journal} {International Journal of Modern Physics D}\
  }\textbf {\bibinfo {volume} {21}},\ \bibinfo {eid} {1242001} (\bibinfo {year}
  {2012}{\natexlab{b}})},\ \Eprint {http://arxiv.org/abs/1205.3160}
  {arXiv:1205.3160 [astro-ph.HE]} \BibitemShut {NoStop}%
\bibitem [{\citenamefont {{Das}}\ and\ \citenamefont
  {{Mukhopadhyay}}(2013{\natexlab{a}})}]{das13c}%
  \BibitemOpen
  \bibfield  {author} {\bibinfo {author} {\bibfnamefont {U.}~\bibnamefont
  {{Das}}}\ and\ \bibinfo {author} {\bibfnamefont {B.}~\bibnamefont
  {{Mukhopadhyay}}},\ }\href@noop {} {\bibfield  {journal} {\bibinfo  {journal}
  {ArXiv e-prints}\ } (\bibinfo {year} {2013}{\natexlab{a}})},\ \Eprint
  {http://arxiv.org/abs/1304.3022} {arXiv:1304.3022 [astro-ph.SR]} \BibitemShut
  {NoStop}%
\bibitem [{\citenamefont {{Nityananda}}\ and\ \citenamefont
  {{Konar}}(2013)}]{nitya13}%
  \BibitemOpen
  \bibfield  {author} {\bibinfo {author} {\bibfnamefont {R.}~\bibnamefont
  {{Nityananda}}}\ and\ \bibinfo {author} {\bibfnamefont {S.}~\bibnamefont
  {{Konar}}},\ }\href@noop {} {\bibfield  {journal} {\bibinfo  {journal} {ArXiv
  e-prints}\ } (\bibinfo {year} {2013})},\ \Eprint
  {http://arxiv.org/abs/1306.1625} {arXiv:1306.1625 [astro-ph.SR]} \BibitemShut
  {NoStop}%
\bibitem [{\citenamefont {{Chamel}}\ \emph {et~al.}(2013)\citenamefont
  {{Chamel}}, \citenamefont {{Fantina}},\ and\ \citenamefont
  {{Davis}}}]{chame13}%
  \BibitemOpen
  \bibfield  {author} {\bibinfo {author} {\bibfnamefont {N.}~\bibnamefont
  {{Chamel}}}, \bibinfo {author} {\bibfnamefont {A.~F.}\ \bibnamefont
  {{Fantina}}}, \ and\ \bibinfo {author} {\bibfnamefont {P.~J.}\ \bibnamefont
  {{Davis}}},\ }\href@noop {} {\bibfield  {journal} {\bibinfo  {journal} {ArXiv
  e-prints}\ } (\bibinfo {year} {2013})},\ \Eprint
  {http://arxiv.org/abs/1306.3444} {arXiv:1306.3444 [astro-ph.SR]} \BibitemShut
  {NoStop}%
\bibitem [{\citenamefont {{Coelho}}\ \emph {et~al.}(2013)\citenamefont
  {{Coelho}}, \citenamefont {{Marinho}}, \citenamefont {{Malheiro}},
  \citenamefont {{Negreiros}}, \citenamefont {{C{\'a}ceres}}, \citenamefont
  {{Rueda}},\ and\ \citenamefont {{Ruffini}}}]{coelh13}%
  \BibitemOpen
  \bibfield  {author} {\bibinfo {author} {\bibfnamefont {J.~G.}\ \bibnamefont
  {{Coelho}}}, \bibinfo {author} {\bibfnamefont {R.~M.}\ \bibnamefont
  {{Marinho}}, \bibfnamefont {Jr.}}, \bibinfo {author} {\bibfnamefont
  {M.}~\bibnamefont {{Malheiro}}}, \bibinfo {author} {\bibfnamefont
  {R.}~\bibnamefont {{Negreiros}}}, \bibinfo {author} {\bibfnamefont {D.~L.}\
  \bibnamefont {{C{\'a}ceres}}}, \bibinfo {author} {\bibfnamefont {J.~A.}\
  \bibnamefont {{Rueda}}}, \ and\ \bibinfo {author} {\bibfnamefont
  {R.}~\bibnamefont {{Ruffini}}},\ }\href@noop {} {\bibfield  {journal}
  {\bibinfo  {journal} {ArXiv e-prints}\ } (\bibinfo {year} {2013})},\ \Eprint
  {http://arxiv.org/abs/1306.4658} {arXiv:1306.4658 [astro-ph.SR]} \BibitemShut
  {NoStop}%
\bibitem [{\citenamefont {{Bera}}\ and\ \citenamefont
  {{Bhattacharya}}(2014)}]{bera14}%
  \BibitemOpen
  \bibfield  {author} {\bibinfo {author} {\bibfnamefont {P.}~\bibnamefont
  {{Bera}}}\ and\ \bibinfo {author} {\bibfnamefont {D.}~\bibnamefont
  {{Bhattacharya}}},\ }\href@noop {} {\bibfield  {journal} {\bibinfo  {journal}
  {ArXiv e-prints}\ } (\bibinfo {year} {2014})},\ \Eprint
  {http://arxiv.org/abs/1405.2282} {arXiv:1405.2282 [astro-ph.SR]} \BibitemShut
  {NoStop}%
\bibitem [{\citenamefont {{Das}}\ and\ \citenamefont
  {{Mukhopadhyay}}(2013{\natexlab{b}})}]{das13a}%
  \BibitemOpen
  \bibfield  {author} {\bibinfo {author} {\bibfnamefont {U.}~\bibnamefont
  {{Das}}}\ and\ \bibinfo {author} {\bibfnamefont {B.}~\bibnamefont
  {{Mukhopadhyay}}},\ }\href {\doibase 10.1103/PhysRevLett.110.071102}
  {\bibfield  {journal} {\bibinfo  {journal} {Physical Review Letters}\
  }\textbf {\bibinfo {volume} {110}},\ \bibinfo {eid} {071102} (\bibinfo {year}
  {2013}{\natexlab{b}})},\ \Eprint {http://arxiv.org/abs/1301.5965}
  {arXiv:1301.5965 [astro-ph.SR]} \BibitemShut {NoStop}%
\bibitem [{\citenamefont {{Das}}\ \emph {et~al.}(2013)\citenamefont {{Das}},
  \citenamefont {{Mukhopadhyay}},\ and\ \citenamefont {{Rao}}}]{das13b}%
  \BibitemOpen
  \bibfield  {author} {\bibinfo {author} {\bibfnamefont {U.}~\bibnamefont
  {{Das}}}, \bibinfo {author} {\bibfnamefont {B.}~\bibnamefont
  {{Mukhopadhyay}}}, \ and\ \bibinfo {author} {\bibfnamefont {A.~R.}\
  \bibnamefont {{Rao}}},\ }\href {\doibase 10.1088/2041-8205/767/1/L14}
  {\bibfield  {journal} {\bibinfo  {journal} {\apjl}\ }\textbf {\bibinfo
  {volume} {767}},\ \bibinfo {eid} {L14} (\bibinfo {year} {2013})},\ \Eprint
  {http://arxiv.org/abs/1303.4298} {arXiv:1303.4298 [astro-ph.HE]} \BibitemShut
  {NoStop}%
\end{thebibliography}%

\end{document}